\documentclass[showpacs,aps]{revtex4}
\usepackage{epsfig}
\begin{document}

\title{  
      Adiabatic nucleus-nucleus potential at near-barrier energies from selfconsistent calculations   
         }

\author{Janusz Skalski  } 
\affiliation{
A. So\l tan Institute for Nuclear Studies,\\
ul. Ho\.za 69, PL- 00 681, Warsaw, Poland }
 \email{ jskalski@fuw.edu.pl
 }

\date{ \today}

\begin{abstract}
 Adiabatic fusion potentials, extended well into compound nucleus region, are 
 calculated for a number of reactions within the static 
 Hartree-Fock method with the Skyrme SkM* interaction. 
 The calculated fusion barriers agree quite well with the data, in spite of 
  considerable errors in reaction $Q$ values. This suggests some error 
 cancellation, possibly with the relative kinetic energy term. Calculated barrier  
 heights are consistent with the idea of fusion hindrance in tip collisions. 
 Adiabatic potentials differ considerably from the results of the frozen density approximation. 
 We briefly discuss relative positions and heights of the fusion and fission potentials    
  and speculate on their possible connection with the fusion hindrance for large $Z_TZ_P$.    

\end{abstract}

\pacs{25.70.Jj,21.60.Jz }

\maketitle


\section{ Introduction}

  Entrance channel potential is one of the principal ingredients in 
 the description of heavy ion collisions. 
 It controls not only the capture probability, but is also  
 related to a substantial hindrance of the compound nucleus (CN) formation, 
 observed experimentally as a large probability of quasifission in reactions 
 between targets and projectiles with $Z_TZ_P>$1800. 
 In this way it is crucial for prospects of the synthesis of the heaviest 
 elements. 

 Up to now, calculations of the entrance channel potentials were performed mostly within some 
 phenomenological models. The mean field Hartree-Fock (HF) methods were applied only with the assumption 
 of frozen densities (see \cite{DN,Dobr} for recent examples), which involves approximations 
 spoiling selfconsistency. The same assumption underlies the nuclear matter approach, see e.g. \cite{JD}.  
 It was sometimes argued that such instantaneous, diabatic fusion barrier should be expected in 
 actual heavy ion collisions due to the short time scale involved in passing 
 over the barrier. Nevertheless, too high barriers and unrealistic, strongly repulsive potentials 
 obtained in the frozen density regime \cite{DN,Dobr} point to a limited relevance of this method. 
 We think that the adiabatic potential is a necessary ingredient in the selfconsistent study of nuclear 
 fusion in much the same way as the static barrier is a necessary first step in a study of nuclear fission.  
 
 In this work, we have performed HF calculations for a number of target-projectile 
 combinations and found the adiabatic nucleus-nucleus potential and (outer) 
 fusion barrier. As might be expected, the adjustment of the nuclear matter to the 
 mutual interaction of target and projectile gives markedly different results from those of the 
 frozen density approach. 

 Within the mean field, for a deformed target or projectile, there are  
 many fusion barriers, depending on the relative orientation of fragments. 
 We have calculated potentials for tip and side collisions, i.e. for 
 the configuration with the symmetry axis of the deformed fragment parallel and 
 perpendicular to the relative distance vector. Effects of the fragment orientation 
 and deformation on fusion were proposed in \cite{MI}.  
 Quite recently, experimental evidence was presented \cite{UO,SmNi,NdGe} showing that 
 at least part of the CN formation hindrance is related to the dominance of quasifission 
 in tip collisions. 

 We have used mostly the Skyrme SkM* interaction \cite{Bartel},   
 originally invented to properly fit the fission barriers. We are aware of previous  
 HF calculations of the potential energy in a dinuclear regime, like e.g. \cite{Gogny}
 (see also recent work \cite{Warda}), however, their objective was nuclear scission, where 
 deformations of separating fragments much differ from those expected in the fusion entrance channel. 

 The aim of this paper is twofold. First, we provide some systematics of the 
 selfconsistent fusion barriers and check against the experimental data  
 and the recent calculations within the frozen density regime 
 \cite{DN} with the same interaction. 
 In our study, we include some systems with well measured fusion cross sections as well as 
 reactions on $^{208}$Pb target of the type studied in GSI Darmstadt 
 \cite{110,HM} and $^{48}$Ca-induced reactions used
 in recent experiments at JINR in Dubna \cite{112,114,116},    
  both reported as leading to the synthesis of the heaviest elements.  
 Second, we look for a relative position (deformations) and height of the fusion channels with 
 respect to the fission valley to figure out quantum transitions implied by the fusion process. 
 Our original motive was to find whether the adiabatic energy surface offers some clue to the 
 fusion hindrance.  

 Some preliminary results of this study have been published in \cite{hfra}.
 Differences between the present barriers and those in \cite{hfra} 
 come from more extensive HF iterations and the improved numeric treatment of kinetic energy.  


\section{ Method of calculations}

 We calculate static potentials, which means that we neglect 
 currents and other time-odd quantities in the Skyrme energy 
 functional and consider time-reversal invariant HF states. This seems 
  reasonable at energies close to the Coulomb barrier. 
  The potential between nuclei 1 and 2 in the central collision is then  
 \begin{equation}
 \label{form}
  V(R)=E(R)+B_1+B_2,
 \end{equation}
 where $E(R)$ is the (negative) HF energy of a dinuclear complex at the 
 distance $R$ and $B_i$ are the (positive) binding energies of target and 
 projectile. In order to have a numerically consistent treatment, $B_i$, $i=1,2$, 
 and $E(R)$ are calculated with the same HF code. 
 In the present calculation pairing is neglected.

 The HF equations have been solved on a spatial mesh of a size proper to the colliding system.  
 Our code assumes two plane symmetries, i.e. it allows for the mass asymmetry along one direction. 
 We consider tip and side collisions, with the angle between the symmetry axis of a 
 deformed nucleus and the line connecting centers of two fragments equal to 
 0$^o$ and 90$^o$. 

 Initially, two sets of wave functions corresponding to two fragments are placed at 
 a chosen distance being an integer multiple of the mesh spacing (0.69 or 0.77 fm). 
 The HF equations are solved by the imaginary-time evolution. Wave functions are 
 kept orthonormal and this enforces the Pauli principle. For fragments 
 placed close enough, the necessary rearrangement of orbitals occurs already at 
 the beginning of the HF procedure and avoids higher than normal densities. 
 Final wave functions correspond to the local minima of the energy functional 
 to which the initial configuration converged. For compact 
 configurations of the CN, i.e. small $R$, there may be quite a few such minima, 
 corresponding to different valleys in the nuclear energy (hyper-)surface. 
 Thus, in contrast to the fusion barrier, 
 the entrance channel (fusion) potential $V(R)$ is not unique.
  The distance $R$ between two fragments is calculated as the distance between the 
 mass centers of two half-spaces containing $A_1$ and $A_2$ nucleons.

  Since HF states are superpositions of various total momentum eigenstates,
  one has to include the center of mass (cm) kinetic energy correction. 
  The expectation value of the cm kinetic energy for an $N$-fermion Slater 
  determinant reads 
  \begin{equation}
  \label{HFK}
  E_{cm}(N)=\frac{1}{2MN}(\sum_{k=1}^N \langle k\mid{\bf p}^2\mid k\rangle
  -\sum_{k\ne l}^N \mid\langle k\mid{\bf p}\mid l\rangle\mid^2),
 \end{equation}
 with $k,l$ labelling occupied single particle states. Only exchange two-body 
 terms are present in Eq.(\ref{HFK}) as the diagonal matrix elements of momentum 
 vanish in the time-reversal invariant states. 
 In most of the HF calculations two-body terms in $E_{cm}$ are neglected (for a
 competent review of the cm corrections in HF see \cite{BRRM}). 
 With the Skyrme SkM* force, $E_{cm}$ is taken as the average kinetic energy, 
 $\langle t \rangle=\sum_{k - occ}\langle k \mid {\hat t}\mid k \rangle /A$. 
 
 Independently of the approximations made, for a Slater determinant describing two separated 
 fragments with $N_1$ and $N_2$ nucleon wave functions, $N_1+N_2=N$, 
 with vanishing mutual momentum overlaps, $\langle k_1\mid{\bf p}\mid l_2\rangle=0$,
 one obtains the relation
  \begin{equation}
  \label{as}
  E_{cm}(N_1)+E_{cm}(N_2)=E_{cm}(N)+\frac{N_2 E_{cm}(N_1)+N_1 E_{cm}(N_2)}{N}.
  \end{equation}
 This implies that in any HF calculation which includes $E_{cm}$, the nucleus-
 nucleus potential $V(R)$ of Eq.(\ref{form}) will tend to the value of the 
 second term on the right hand side of Eq.(\ref{as}), when $R$ tends to 
 infinity. This term is just the expectation value of the kinetic energy of 
 the relative motion of the two fragments. 
 In order to preserve the usual meaning of the Coulomb barrier 
 one has to subtract this asymptotic term in Eq.(\ref{form}). 
 Specifically, with the SkM* force and for fragments with $A_1$ and $A_2$ nucleons, 
 $A_1+A_2=A$, one has to subtract 
 $(A_2 \langle t_1 \rangle +A_1 \langle t_2 \rangle)/A$. 

 This subtraction has been included in the frozen density calculations \cite{DN,Dobr}.
 However, it is surely incorrect when two fragments become one, i.e., 
 for compact configurations of the system. 
 As we do not know how kinetic energy of the relative motion of the fragments 
 transforms into potential energy of the combined system, and still want 
 to compare with the existing frozen density results, we also calculate $V(R)$ normalized 
 to zero at infinity. One has to keep in mind that such potential underestimates  
 the true $V(R)$: the smaller the distance $R$, the larger the offset which reaches 
 $-14$ to $-18$ MeV at the CN ground state.   
 The product $Z_TZ_P$ controls the overlap of target and projectile at the fusion barrier: 
 the larger $Z_TZ_P$, the larger the overlap. Hence, one can expect that $E_{cm}$ correction, 
 valid for separated fragments, produces too low fusion barriers for fragments with 
 large $Z_TZ_P$. 

\section{ Results and discussion}

 Examples of entrance channel potentials are shown in Figs. 1,2.     
 In Fig. 1 we show the tip- and side collision barrier for the $^{244}$Pu$+^{48}$Ca reaction. 
 Two results are shown, differing by the number of performed HF iterations.    
 Adiabatic (lower) barriers were obtained after three long series of iterations, required in view of 
 an inefficient convergence of the pure HF without pairing. Only one series of iterations produced 
  the other (higher) plotted barriers. 
 The quantities obtained in this way were in fact published in \cite{hfra}. 
It is seen that an incomplete convergence in the true HF mimics to some extent the frozen density results
 \cite{DN,JD}: for systems with large $Z_TZ_P$ there is a potential pocket 
 inside the fusion barrier and $V(R)$ rises steeply with decreasing $R$. 
 In the frozen density method this results from a higher than normal density in the 
  overlap region. In the HF, it is the initial orthogonalization which  
   disturbs density in the overlap region and rises energy. A subsequent quenching is necessary 
   to lose this excitation. 

In quantitative terms, even non-fully adiabatic selfconsistent potentials 
are lower and rise much less steeply with decreasing $R$ than the frozen density results. 
 This rise vanishes in fully adiabatic potentials, as seen in Figs. 1,2. These potentials 
 decrease with decreasing $R$, except for the superheavy systems, where the Coulomb barrier becomes 
 a local plateau at the level of, or even below, the ground state of CN. In addition to 
 the $^{208}$Pb$+^{70}$Zn reaction, potentials with a plateau in the place of the barrier are obtained 
 also for the $^{208}$Pb$+^{82}$Ge and $^{132}$Sn$+^{132}$Sn systems. A comparison of fusion barrier 
 heights shows that for reactions on $^{208}$Pb and actinide targets leading to very heavy nuclei 
 the adiabatic barriers are by 10-20 MeV lower than the frozen 
 density ones \cite{DN} (see also Table 1).
 Taken at the face value, the adiabatic results suggest that, after passing the 
 fusion barrier, systems with sufficiently low $Z_PZ_T$ experience a steady 
 drivig force towards CN formation, while with increasing $Z_TZ_P$ this force weakens and even 
 changes to repulsion. 
 
 It may be noticed in Fig. 1 for the reaction $^{244}$Pu$+^{48}$Ca that  
 the shorter and higher side collision potential offers larger driving force towards CN 
 than the longer but lower tip collision one. 
 The difference in the barrier position for tip- and side collision for  
 $^{244}$Pu+$^{48}$Ca system, $R_{tip}-R_{side}$, is about 2 fm. Similar situation occurs for 
  the reaction $^{110}$Pd$+^{110}$Pd (Fig. 2), for which
 the tip-tip collision potential shows a plateau with fluctuations.  
 On the other hand, the higher side-side collision barrier gives chance to 
 a quite steep (20 MeV) slide towards CN pocket (see below).  
 The potential for the $^{208}$Pb$+^{70}$Zn 
  reaction, on the other hand, does not provide any force in the CN direction. 
  This could suggest that the entrance channel potentials give a better chance 
  of fusion in the side collisions with prolate deformed actinide targets 
  as compared to reactions on $^{208}$Pb target. 

 When comparing various results for barriers one has to remember that 
  with rising $Z_TZ_P$ 1) the frozen density approach 
 becoms poorer, 2) the correction for cm energy becomes questionable and may  
  produce too low barriers, 3) the adiabaticity effect on the lowering of the barrier 
  increases.  
 
 Calculated fusion barriers $B_{cal}$, taken as the locally highest value of $V(R)$, 
 rounded to 0.5 MeV, are compared in Table 1 to the Bass interaction barriers \cite{Bass} 
 and to the recently given threshold barriers $B_{thre}$ \cite{Wil}. 
 The interaction barriers are chosen here instead of the fusion barriers 
 \cite{Bass} since they  much better approximate the experimental barriers 
 for heavy systems. The threshold barriers \cite{Wil} are derived from 
 sufficiently detailed fusion data and should correspond to the calculated 
 adiabatic barriers. The values of $B_{thre}$ for the heaviest systems are 
 based on the capture data \cite{Itkis}.  
 For deformed targets, both the calculated tip and side (in parentheses) 
 collision barriers are given.
 In order to appreciate excitation of a compound system at the fusion barrier 
 we also give in Table 1 the experimental (or extrapolated) 
 reaction $Q$ values. 

 Out of nine threshold barriers, two are overestimated and six underestimated in the calculations by  
 up to 4 MeV (it seems that $B_{thre}$ for the $^{238}$U$+^{16}$O reaction corresponds to 
 tip collision \cite{UO}). One, for the $^{238}$U$+^{48}$Ca reaction, is 6 MeV below the 
 calculated side collision barrier and 11 MeV above the tip collision one.
  Although for two reactions leading to $^{214}$Th CN the barriers $B_{thre}$ 
  are not known exactly, their rough estimate from data \cite{SmNi} is possible.
  For the reaction $^{182}$W$+^{32}$S the estimate agrees well with our $B_{cal}$, 
  while for $^{154}$Sm$+^{60}$Ni it seems close to the calculated tip collision 
  barrier, although the lack of the low energy data causes large uncertainty. 
  It should be stressed that the lack of observed evaporation residues (ERs) at $E_{cm}$=175 
  and 182 MeV in the experiment \cite{SmNi} served as an argument for the fusion hindrance in 
  tip collisions.  
  
 The fusion barrier calculated for the recently measured reaction 
 $^{132}$Sn$+^{64}$Ni is higher than $B_{thre}$ suggested by the data \cite{SnNi}, but 
 lower than the Bass interaction barrier assumed in the coupled channels 
 analysis there. 
 The $^{110}$Pd nucleus, previously considered vibrational (see e.g. \cite{PdS}), 
 is now well established by the Coulomb excitation studies \cite{pddef} 
 as a rotor with deformation $\beta_2\approx$ 0.25.  
 We have obtained a shallow prolate-deformed HF minimum at $Q=$7.4 b, 
 consistent with this, and used it to calculate tip-tip and side-side collision 
 barriers given in Fig. 2 and Table 1. 
 The $^{110}$Pd$+^{110}$Pd$\rightarrow ^{220}$U reaction is a very well known 
 case of the severe fusion hindrance \cite{Pd110}, at least by four orders of magnitude.   
 The lowest measured point at $E_{cm}=$228 MeV is already above the calculated side collision 
 barrier.   
 
 For most reactions leading to superheavy nuclei there are too few data to extract 
 threshold barriers, even from capture data. However, the  
 data on evaporation residue formation give some upper bound on the height of the 
 fusion barrier, as their production at $E_{cm}$ below the lowest barrier is improbable. 
 For $^{208}$Pb target, the $^{271}110$ ERs were detected 
 following reaction with $^{64}$Ni at $E_{cm}=235.3-238.2$ MeV \cite{110,110B}, 
 while in reaction with $^{70}$Zn two observed $^{277}112$ ERs were produced 
 at $E_{cm}=$257.2 and 259.1 MeV \cite{HM}. In reactions with $^{48}$Ca 
 projectiles on actinide targets, two ERs were observed in the reaction on 
 $^{238}$U at $E_{cm}=190.5-193.9$ MeV \cite{Og2}, three events in the reaction on 
 $^{244}$Pu target at $E_{cm}=194.5-202$ MeV 
 \cite{114,Og1} and one event for $^{248}$Cm target at $E_{cm}=199.7-205.1$ MeV \cite{116}.
 All these bombardment energies in the c.m. system are higher than our calculated barriers: 
 by at least 6 and 13 MeV for $^{64}$Ni and $^{70}$Zn projectiles on $^{208}$Pb target, 
 and by 2-3 MeV higher than the calculated side collision barriers for actinide targets. 

 Overall reasonable agreement of the calculated fusion barriers with data 
 is quite astonishing in view of the deficiency shown by the SkM* force in reproducing 
 the experimental binding energies. 
 As follows from our calculations and the recently calculated mass table 
 \cite{SkM*mas} there is a large underestimation of binding for 
 heavy and superheavy nuclei with the SkM* force, e.g. about 9 MeV for 
 $^{182}$W and $^{244}$Pu, but already about 14 MeV for $^{256}$No, 
 and expected 18 MeV for $^{292}114$, according to usually fair 
 estimates of the phenomenological Thomas-Fermi model \cite{MS}. Moreover, there are some errors 
  specific to SkM* force, like too large bindings of 
 $^{48}$Ca (by 4 MeV) and $^{132}$Sn (by 7.5 MeV). 
 
 It is reasonable to ask about a correlation between 
 errors in fusion barriers and those in reaction $Q$ values.  
 It turns out that the $Q$ values are overestimated by up to 5 MeV 
  for lighter compound systems, i.e. similar in magnitude but mostly opposite to 
   the errors in barriers. 
 However, for heavier compound systems the errors in barriers are much smaller 
 than those in $Q$ values, e.g. $Q$ values are too 
 large by about 11 MeV for $^{90}$Zr$+^{90}$Zr, about 17 MeV for 
 $^{48}$Ca$+^{208}$Pb, expected about 13 MeV too large for $^{48}$Ca$+^{244}$Pu.  
  All this points to some error cancellation, which, as discussed in the previous section, 
  might be related to the overestimated relative kinetic energy correction.  

 An intriguing question is whether the calculated potentials $V(R)$ may be related 
 quantitatively to the fusion hindrance seen in experiment. In \cite{DN}, a sufficiently 
 deep pocket in the frozen density potential $V(R)$ was proposed as a necessary 
 precondition for fusion. The quantification of this assertion is difficult, as the depth 
 and the very existence of the pocket relies on the unrealistic frozen density intrinsic 
 barrier, which exists e.g. already for $^{90}$Zr+$^{90}$Zr and $^{208}$Pb+$^{48}$Ca reactions, 
 for which no fusion hindrance is experimentally observed \cite{Itkis,zr90}. 

 Reasoning intuitively, a relative situation and heights of fusion and fission valleys 
 should be relevant for prospects of fusion: to be captured in the CN configuration, 
  a system must pass inside the CN fission barrier.     
  In order to gain some orientation in this respect, a more detailed calculations have been 
  performed for a few nuclei. The results for $^{292}$114 ($^{244}$Pu$+^{48}$Ca reaction) 
  are shown in Fig. 3. When showing fission and fusion barriers in one figure 
 the kinetic energy of the fragment relative motion {\it has not been subtracted}
 from the fusion barrier. Instead of the distance $R$ between two fragments the 
 quadrupole moment $Q_{20}$ of the compound system is shown on the abscissa. 
 The calculated mass- and axially symmetric fission barrier of 15 MeV is heavily 
 overestimated as compared to the experimental estimate  $B_{fis}\approx 6.5$ MeV \cite{fisbar}, 
  consistent with the Strutinsky method calculations \cite{Smol}. 
 This comes partly from the lack of pairing and neglect of nonaxiality. Accounting for both 
 these effects still leaves a few MeV discrepancy between the data and the barrier obtained 
  with the SkM* force. Similarly, the second hump of the barrier will be lowered by pairing and 
  mass-asymmetry. Therefore, Fig. 3 gives a general idea about relative position of 
 important points of the nuclear energy landscape rather than exact barrier heights. 

  The tip-collision potential lies below the second hump of the fission barrier and 
  corresponds to a sizable mass asymetry, increasing with $Q_{20}$. 
  In the right portion of the figure, we extend fission barrier to 
  mass asymmetry by showing cuts through the potential energy surface at fixed $Q_{20}=$100, 
  125 and 140 b vs. octupole moment $Q_{30}$. 
  It is seen that the mass-asymmetric minimum becomes 
  energetically well separated from the symmetric fission valley with increasing $Q_{20}$. 
  The nuclear states building fusion potential are very close to the states in the fission valley
  with alike quadrupole and octupole moments, e.g. 
  at $Q_{20}=140$ b and  $Q_{30}\approx 25\cdot 10^3$ fm$^3$, at $Q_{20}\approx 125$ b and 
  $Q_{30}\approx 20\cdot 10^3$ fm$^3$ and at $Q_{20}\approx 100$ b and $Q_{30}\approx 
  12\cdot 10^3$ fm$^3$. Slight differences in energy (1-2 MeV) come from differences in necking.    
  One can say that the tip collission potential arises from traversing potential 
  energy surface, in this case not quite along the conditional minima at fixed $Q_{20}$, as 
  these lie slightly lower.  

  This overall picture suggests that, after surpassing the fusion barrier, the system has a good 
  chance to fall inside the outer fission barrier. If some energy is lost to excitation on the way,  
  it may be stopped at the higher inner barrier. 
  One may expect that fusion in the tip collisions is hindered in this case. 
    
The side collision potential lies above both humps of the fission barrier and, after surpassing 
the fusion barrier, the system can roll down all the way to the CN ground state, even if some 
energy is transferred to excitation. (Looking at Fig. 3 it is good to remember that the side 
collision barrier has a large axial asymmetry.) 
 
   For $^{256}$No ($^{208}$Pb$+^{48}$Ca reaction, not shown) the mass-asymmetric entrance 
   channel potential leads inside the higher, inner hump of the mass-symmetric fission barrier. 
   The calculated height of the latter,  15 MeV, again exceeds the experimental estimate 
   consistent with the value $\approx$8 MeV obtained in the Strutinsky-like 
   studies, but is similar to the barrier (12.6 MeV) calculated for $^{254}$No 
   with SkLy4 force \cite{Dug}. 
    The lower, secondary hump of the fission barrier does not play any role.  
     Therefore, one should not expect drastic fusion hindrance in this reaction.  

  The above are examples of possible qualitative inferences one can make from 
  knowing adiabatic energy surfaces. 
  Surely, the concept that the relative situation of the fusion potential and fission barriers has 
  relation to fusion hindrance requires further study. It is plausible that not only nuclear 
  configurations, but also configuration changes which occur with some probability, are involved in 
    the resulting loss of the probability of fusion.  

 Conclusions of this work may be stated as follows: 

\noindent
 - Adiabatic fusion barriers calculated with SkM* force compare rather well with experiment, anyway 
 much better than the frozen density results. 

\noindent
 - The barriers compare with experiment much better than the 
   reaction $Q$ values for heavy and superheavy systems. This may suggest that there is some error 
   cancellation: subtraction of a too large relative kinetic energy makes up for too large 
    reaction $Q$ values. 

\noindent
 -  The rise of the potential $V(R)$ with decreasing $R$ seen in the frozen density or not fully adiabatic 
   calculations vanishes in adiabatic potentials, except for the heaviest CN where the fusion barrier top lies 
    below the ground state. Thus, the relation of this rise to the fusion hindrance is not clear.  

\noindent
 -  The comparison of our calculated barrier heights to data supports the idea that in reactions with 
    large $Z_TZ_P$ and deformed fragments fusion is suppressed in tip collisions in spite of the substantially 
    lower fusion barrier. 
    
\noindent
   - Judging from forces driving towards the CN minimum, the formation of superheavy CN may be easier in 
     side collisions with deformed actinides used in JIHR Dubna, than in reactions with spherical targets 
     used in GSI Darmstadt (without considering survivel probability). 
    
    The latter two points suggest that the size of a necessary rearrangement of the nuclear matter 
    drop is more vital for fusion than the barrier heights.  
    One may wonder whether some measure of this rearrangement is not a 
    proper variable to describe the fusion hindrance. One such variable could be $R_{fus}-R_{fis}$, where 
    the position of the inner fission barrier is relevant. 
    Obviously, these speculations call for a more detailed study, including multidimensional energy surfaces
    and pairing.


\newpage

\begin{center}
 
\hspace{5mm}{TABLE 1 - 
Calculated fusion barriers for tip (side) collisions 
in MeV vs. threshold \cite{Wil} and Bass {\it interaction} (not 
 fusion) barriers \cite{Bass}. 
Threshold barrier for $^{90}$Zr$+^{90}$Zr is inferred from 
 \cite{zr90}, that for $^{238}$U$+^{16}$O from \cite{UO}, and for $^{110}$Pd$+^{32}$S from \cite{PdS}. 
 Reaction $Q$ values (experimental/extrapolated \cite{Audi}, or predicted (with asterisk) \cite{MS}) 
 are given in column 2.} 

\vspace{1mm}

\begin{tabular}{c|cccc} \hline
 System  &  $Q$  & $B_{cal}$ & $B_{thre}$& $B^{int}_{Bass}$   \\
 \hline

$^{40}$Ca+$^{40}$Ca & 14.3   & 52  &  50.2$\pm$0.2  &  52.5   \\

$^{64}$Ni+$^{64}$Ni & 48.8   & 86.5 &  89.5$\pm$0.3  &  94.5   \\

$^{90}$Zr+$^{40}$Ca & 57.3 & 92    &  92.7$\pm$0.6  &  97.7    \\

$^{96}$Zr+$^{40}$Ca & 41.1  & 85  &  87.5$\pm$0.3  & 96.6    \\

$^{90}$Zr+$^{90}$Zr & 157.3 & 175   & $\sim$175.85 & 181.0  \\

$^{110}$Pd$+^{32}$S & 35.4 & 78(79.5)  & 80.4$\pm$0.2 & 88.9   \\

$^{110}$Pd+$^{110}$Pd & 199.7 & 200(222.5)   & - & 228.3  \\

$^{182}$W$+^{32}$S & 84.9  & 121.5(124) & -  & 134.3  \\

$^{154}$Sm$+^{60}$Ni & 147.6 &  170(185.5) & -  & 191.5 \\

$^{238}$U+$^{16}$O & 38.3 & 67(71) & $\sim$71  & 82.9 \\

$^{132}$Sn+$^{64}$Ni & 111.1 & 147 &  -  &  155.7  \\

$^{132}$Sn+$^{132}$Sn & 260.8 $^*$ & 254.5  & - & 257.4  \\
 \hline

$^{208}$Pb+$^{48}$Ca & 153.8  & 170.5 &  169$\pm$2 & 176.1 \\

$^{208}$Pb+$^{64}$Ni & 225.1 & 229   & - & 241.3 \\

$^{208}$Pb+$^{70}$Zn & 244.9 $^*$ & 244  & - & 256.3 \\

$^{208}$Pb+$^{82}$Ge & 262.5 $^*$ & 260.5 & - & 268.8 \\
 \hline

$^{238}$U+$^{48}$Ca & 160.8 $^*$ & 171(188)  & 182$\pm$2 &  193.8 \\

$^{244}$Pu+$^{48}$Ca & 163.0 $^*$ & 174(192) & - & 197.3  \\

$^{248}$Cm+$^{48}$Ca & 169.3 $^*$ & 183(196.5)& -  & 201.0  \\

$^{250}$Cf+$^{48}$Ca & 177.0 $^*$ & 187.5(201) & -  & 205.1  \\
 \hline
\end{tabular}

\end{center}

\newpage

\begin{figure}[t]
\begin{minipage}[t]{60mm}
\centerline{\includegraphics[scale=0.3, angle=-90]{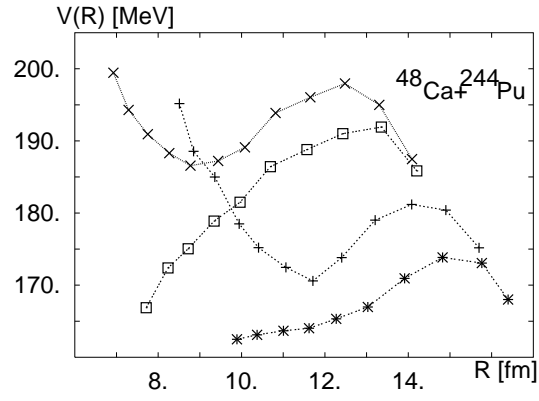}}
\end{minipage}
\caption{{\protect\small Partially and fully adiabatic potentials for $^{244}$Pu$+^{48}$Ca tip 
(pluses, stars) and side (crosses, squares) collision obtained with Skyrme SkM* force. 
  }}
\end{figure}

\begin{figure}[t]
\begin{minipage}[t]{60mm}
\centerline{\includegraphics[scale=0.3, angle=0]{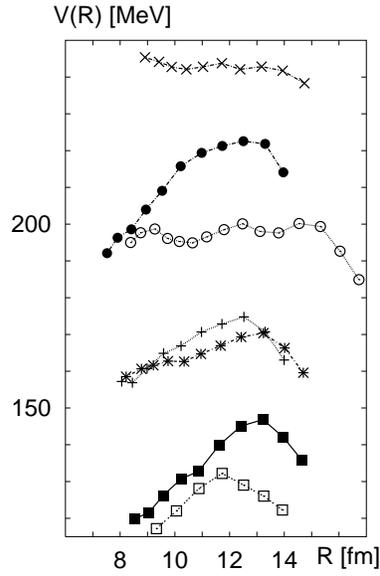}}
\end{minipage}
\caption{{\protect\small Adiabatic nucleus-nucleus potentials for the reactions: $^{90}$Zr$+^{40}$Ca 
(shifted up by 40 MeV, open squares), $^{132}$Sn$+^{64}$Ni (full squares), $^{208}$Pb$+^{48}$Ca (stars), 
$^{90}$Zr$+^{90}$Zr (pluses), $^{110}$Pd$+^{110}$Pd tip-tip (open circles) and side-side collision 
(full circles) and $^{208}$Pb$+^{70}$Zn (crosses). 
 }}
\end{figure}


\begin{figure}[t]
\begin{minipage}[t]{62mm}
\centerline{\includegraphics[scale=0.3, angle=-90]{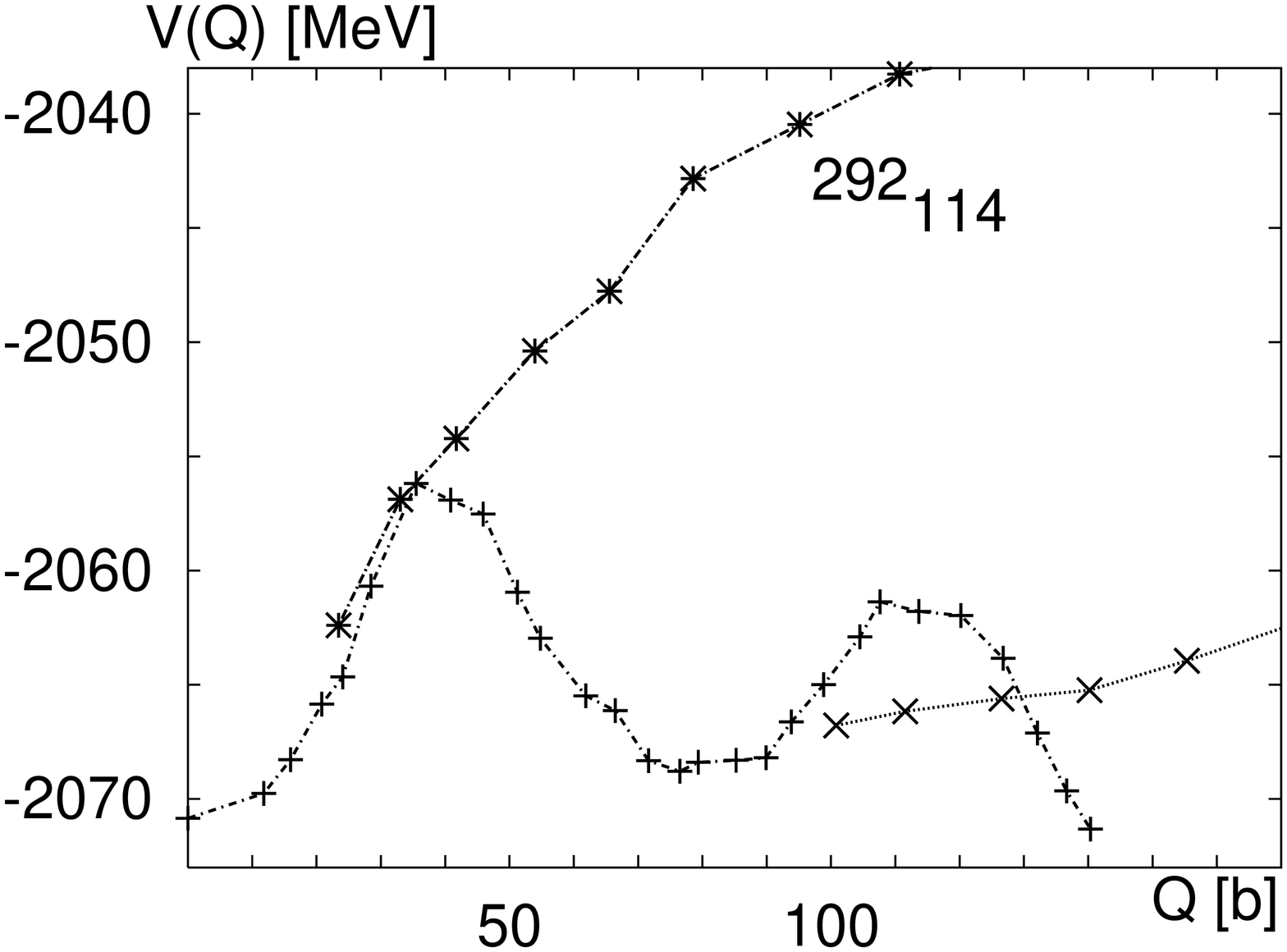}}
\end{minipage}
\hspace{10mm}
\begin{minipage}[t]{62mm}
\centerline{\includegraphics[scale=0.3, angle=-90]{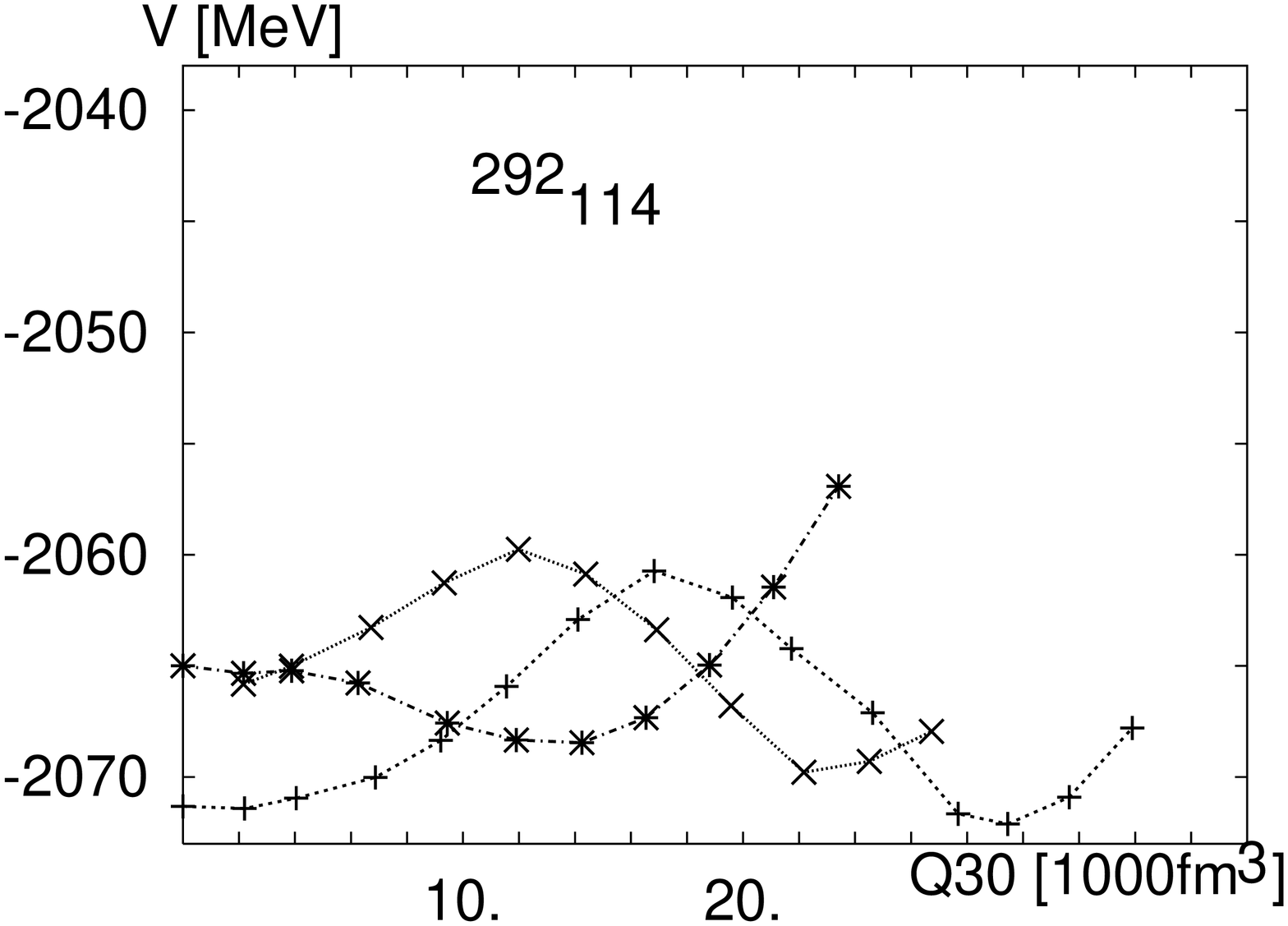}}
\end{minipage}
\caption{{\protect\small Left panel: Mass-symmetric fission barrier (pluses) and fusion barriers for 
the $^{48}$Ca-induced reaction (tip collision - crosses, side collision - stars) for the compound 
system $^{292}$114. Right panel: fission valley in the same system vs. octupole moment $Q_{30}$ for 
$Q_{20}=100$ b (stars), 125 b (crosses) and 140 b (pluses)
. }}
\end{figure}

\end{document}